\documentclass[twocolumn,preprintnumbers,prl,amsmath,amssymb]{revtex4}

\usepackage{graphicx}
\usepackage{bm}

\newcommand{\Sr}{SrFe$_2$As$_2$}
\newcommand{\Ba}{BaFe$_2$As$_2$}
\newcommand{\BaP}{BaFe$_2$As$_{2-x}$P$_x$}
\newcommand{\Ca}{CaFe$_2$As$_2$}
\newcommand{\BaSr}{Ba$_{1-x}$Sr$_x$Fe$_2$As$_2$}
\newcommand{\SrCa}{Sr$_{1-x}$Ca$_x$Fe$_2$As$_2$}
\newcommand{\SrCacc}{Sr$_{0.3}$Ca$_{0.7}$Fe$_2$As$_2$}
\newcommand{\SrCaxray}{Sr$_{0.33}$Ca$_{0.67}$Fe$_2$As$_2$}
\newcommand{\BaSrCa}{(Ba,Sr,Ca)Fe$_2$As$_2$}

\newcommand{\tc}{$T_c$}

\begin{document}

\date{\today}

\title{Uniform chemical pressure effect in solid solutions Ba$_{1-x}$Sr$_x$Fe$_2$As$_2$ and Sr$_{1-x}$Ca$_x$Fe$_2$As$_2$}

\author{S.~R.~Saha, K.~Kirshenbaum, N.~P.~Butch}
\address{Center for Nanophysics and Advanced Materials, Department of
Physics, University of Maryland, College Park, MD 20742}

\author{J. Paglione}
\email{paglione@umd.edu}
\address{Center for Nanophysics and Advanced Materials, Department of
Physics, University of Maryland, College Park, MD 20742}

\author{P. Y. Zavalij}
\address{Department of Chemistry and Biochemistry, University of
Maryland, College Park, MD 20742, USA}

\begin{abstract}

The effect of alkaline earth substitution on structural parameters
was studied in high-quality single crystals of \BaSr\ and \SrCa\
grown by the self-flux method. The results of single-crystal and
powder x-ray diffraction measurements suggest a continuous monotonic
decrease of both $a$- and $c$-axis lattice parameters, the $c/a$
tetragonal ratio, and the unit cell volume with decreasing alkaline
earth atomic radius as expected by Vegard's law. As a result, the
system experiences a continuously increasing chemical pressure
effect in traversing the phase diagram from $x=0$ in \BaSr\ to $x=1$
in \SrCa.

\end{abstract}

\maketitle


The recent discovery of high-temperature superconductivity in iron-based
compounds has attracted much interest. The parent phases of these
compounds generally show antiferromagnetic order
that onsets between 130~K and 200~K,
with superconductivity emerging when the
antiferromagnetic order of the parent compounds is
suppressed~\cite{Kamihara,Cruz,Rotter,Saha,Ren}. This proximity of
magnetic and superconducting order parameters is widely thought to be a
key argument for an unconventional pairing mechanism, likely
mediated by spin fluctuations~\cite{Mazin,Christianson} similar to the cuprates~\cite{Shirane,Keimer}. But in strong
contrast to the copper oxides, superconductivity in iron arsenides
can be induced without changing the carrier concentration, either by
applying external pressure~\cite{Okada,Alireza} or by isovalent chemical substitution.
The highest \tc\ achieved so far in these materials is $\sim
55$~K in SmO$_{1-x}$F$_x$FeAs~\cite{Ren} and
(Sr,Ca)FeAsF~\cite{Zhu,Cheng}. Oxygen-free FeAs-based compounds
with the ThCr$_2$Si$_2$-type (122) structure also exhibit
superconductivity induced by chemical substitution of alkali or transition metal
ions~\cite{Rotter,Sasmal,Sefat,Leithe}, the application of large
pressures~\cite{Alireza,Torikachvili,CaFe2As2,Kumar}, or lattice
strain \cite{saha1},
with transition temperatures as high as $\sim 37$~K.

For the 122 phase, superconductivity has been
induced by substituting Fe with not only 3$d$-transition metals such
as Co and Ni, but also some of the 4$d$- and 5$d$-transition metals.
Recently, Ru, Ir, and Pt substitution for Fe were also shown to induce
superconductivity in \Sr\ and \Ba\ \cite{Schnelle,Han,SahaPt}. Superconductivity with \tc $\sim$ 31~K has also been
shown to occur by  isovalent substitution of P for
As~\cite{Kasahara}. This gives the opportunity to tune magnetic
character without nominally changing charge carrier concentrations, for instance making the interpretation of transport coefficients much simpler
than in the case of charge doping.

In order to investigate the possibility of applying uniform chemical pressure in a continuous manner, we have synthesized the series of solid solutions \BaSr\ and \SrCa\
by substituting isovalent alkaline earth atoms, and investigated the evolution of the crystal structure by high-resolution powder and single-crystal x-ray  diffraction. Here we present our preliminary results that suggest the unit cell of the Ba-Sr-Ca substitution series experiences a monotonic uniform chemical pressure as a function of alkaline earth atomic radius.


Single-crystal samples of \BaSr\ and \SrCa\ were grown using the
FeAs self-flux method~\cite{saha1}. Fe was first separately
pre-reacted with As via solid-state reaction of Fe (99.999\%) powder
with As (99.99\%) powders in a quartz tube of partial atmospheric
pressure of Ar. The precursor materials were mixed with elemental Sr
(99.95\%) with either Ba (99.95\%) or Ca (99.95\%) in the ratio
4:1$-x:x$, placed in an alumina crucible and sealed in a quartz tube
under partial Ar pressure. The mixture was heated to 1150$^\circ$C,
slow-cooled to a lower temperature and then quenched to room
temperature. Typical dimensions of as-grown single crystal specimen
are $\sim$100~$\mu$m thickness and up to 5~mm width. Chemical
analysis was performed using both energy- and wavelength-dispersive
x-ray spectroscopy (EDS and WDS).

\begin{figure}[tbh] \centering
  \resizebox{8cm}{!}{
  \includegraphics[width=8cm]{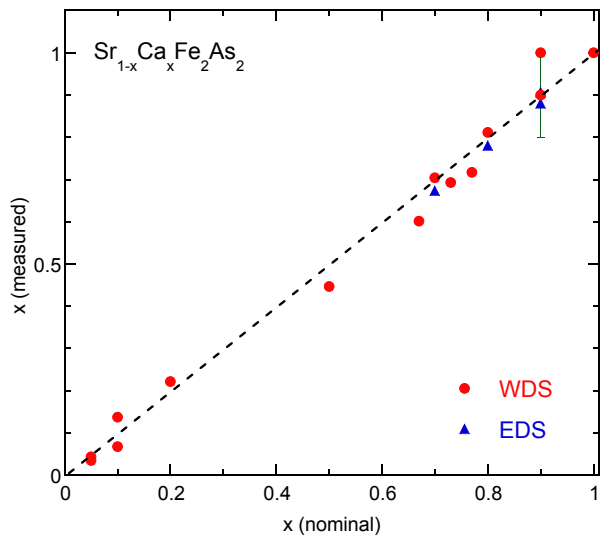}}
  \caption{\label{fig1} Actual Ca concentration of \SrCa\ single-crystal samples as a
function of nominal concentration $x$, as determined by wavelength
dispersive x-ray spectroscopy (data points represent average value
of 8 scanned points for each concentration). Some of the specimens
are also confirmed by energy dispersive X-ray spectroscopy (EDS).
The dotted line is a guide to eye which traces
$x$(measured)=$x$(nominal).}
\end{figure}



Both EDS and WDS analysis of all \BaSr\ and \SrCa\ samples showed the proper 1:2:2
stoichiometry in all specimens reported herein, with no indication of
impurity phases. Figure~\ref{fig1} compares the nominal alkaline earth concentration
$x$ in \SrCa\ crystals with that measured by WDS and EDS analysis, using an average value determined from 8 different
spots on each specimen. As shown by the dotted line guide, the actual
concentrations found by WDS are equal to the nominal
values of $x$ to within experimental error, indicating homogeneous substitution in this series of
solid solutions.

\begin{figure}[tbh] \centering
  \resizebox{8cm}{!}{
  \includegraphics[width=8cm]{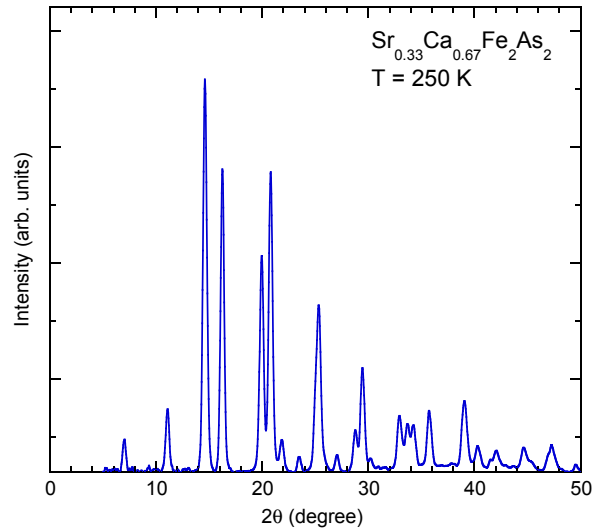}}
  \caption{\label{fig2} Typical x-ray powder diffraction pattern, shown for sample \SrCaxray, obtained by using Mo-K$_{\alpha}$ radiation. The main peaks can be indexed with a tetragonal structure and there are no impurity phases detected within experimental
accuracy.}
\end{figure}

Diffraction patterns were obtained by both powder and single-crystal x-ray diffraction and
Rietfeld refinement (SHELXS-97) to $I4/mmm$ structure. Powder x-ray
diffraction was performed at 250~K using a Smart Apex2 diffractometer
with Mo-K$_{\alpha}$ radiation and a graphite monochromator.
Figure~\ref{fig2} shows a typical x-ray diffraction pattern obtained from a single-crystal sample of \SrCacc. All of the main peaks can be indexed to
the ThCr$_2$Si$_2$ structure, with no impurity phases detected.

\begin{figure}[tbh] \centering
  \resizebox{6cm}{!}{
  \includegraphics[width=6cm]{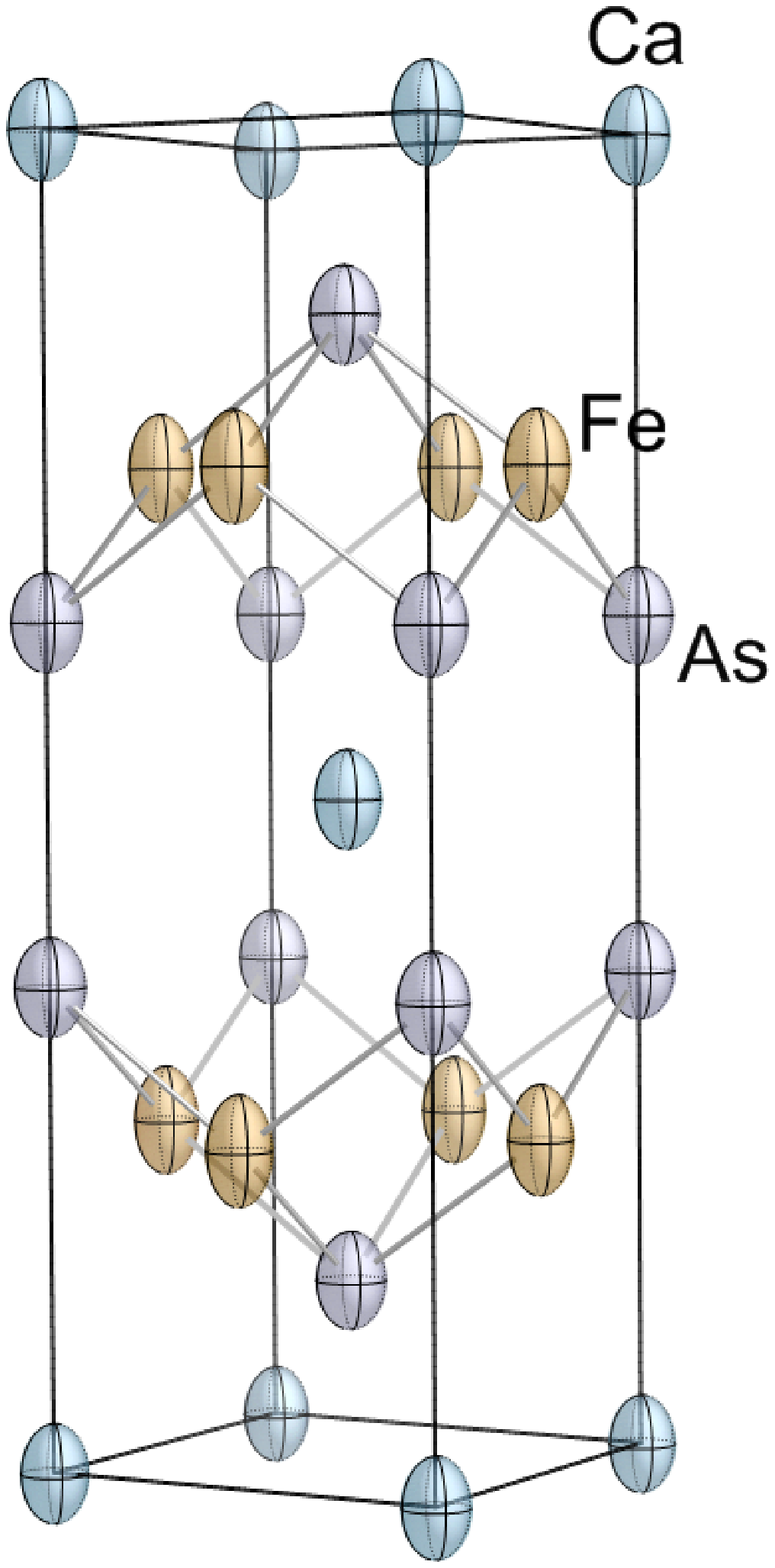}}
  \caption{\label{fig3} Unit cell of \SrCaxray\ determined by single
crystal x-ray diffraction at 250~K.}
\end{figure}

\begin{table}[tbh]

\caption{\label{tabl1}Crystallographic data for \Sr\ and \SrCaxray\
determined by single-crystal x-ray diffraction at 250~K. The
structure was solved and refined using the SHELXS-97 software,
yielding lattice constants with residual factor $R$= 1.36\% and
1.95\% for \Sr\ and \SrCaxray, respectively. }

\footnotesize\rm
\begin{tabular}{lll}
\hline
&\Sr&\SrCaxray\\
\hline
Temperature&250~K&250~K\\
Structure&Tetragonal&Tetragonal\\
Space group&I4/mmm&I4/mmm\\
$a$($\mathrm{\AA}$)&3.9289(3)&3.9066(8)\\
$b$($\mathrm{\AA}$)&=$a$&=$a$\\
$c$($\mathrm{\AA}$)&12.3172(12)&11.988(5)\\
$V$($\mathrm{\AA}^3)$&190.17(4)&182.95(9)\\
$Z$&2&2\\
Density(g/cm$^3$)&6.098 &6.045\\
Atomic parameters:&&\\
Sr/Ca&2$a$(0,0,0)&2$a$(0,0,0)\\
Fe&4d(1/2,0,1/4)&4$d$(1/2,0,1/4)\\
As&4e(0,0,$z$)&4$e$(0,0,$z$)\\
&$z$=0.36035(5)&$z$=0.36423(7)\\
Atomic displacement&&\\
parameters U$_{eq}$ ($\mathrm{\AA}^2$):&&\\
Sr1/Ca1&0.0108(2)&0.0116(5)\\
Fe1&0.0096(2)&0.0125(3)\\
As1&0.00964(17)&0.0119(2)\\
Bond lengths ($\mathrm{\AA}$):&&\\
Sr/Ca-As&3.2677(4) $\times$ 8&3.2062(7) $\times$ 8\\
Fe-As&2.3890(4) $\times$ 4&2.3855(7) $\times$ 4\\
Fe-Fe&2.7782(2) $\times$ 4&2.7624(6) $\times$ 4\\
Bond angles (deg):&&\\
As-Fe-As&110.63(3) $\times$ 2&109.94(4) $\times$ 4\\
 &108.896(14) $\times$ 4&109.24(2) $\times$ 4\\
Fe-As-Fe&71.105(14) $\times$ 4&70.76(2) $\times$ 4\\
\hline

\end{tabular}
\end{table}


\begin{figure}[tbh]
\centering
  \resizebox{8cm}{!}{
  \includegraphics[width=8cm]{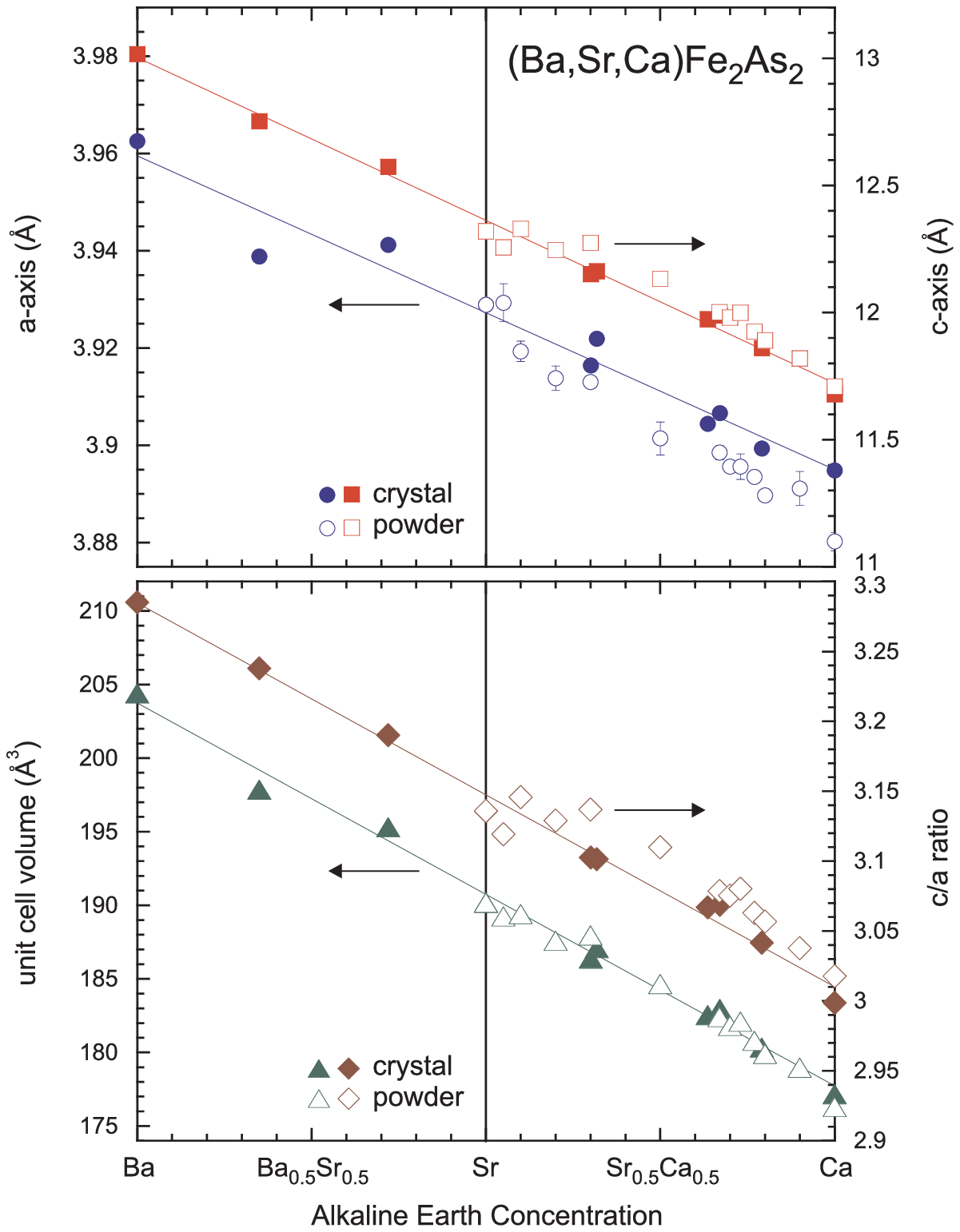}}
  \caption{\label{fig4} Upper panel: Variation of
the $a$- and $c$-axis lattice constants as a function of alkaline
earth substitution in the series \BaSr\ (left half) and \SrCa\
(right half) as determined from single crystal x-ray diffraction
measurements at 250 K of single-crystal samples. Corresponding $c/a$
ratio and unit the cell volume are plotted in the lower panel. In
both panels, solid symbols indicate data acquired using
single-crystal specimens and open symbols represent data determined
by powder x-ray diffraction.}
\end{figure}

Table~1 shows the crystallographic parameters determined by
single-crystal x-ray-diffraction at 250~K in \SrCacc. A Bruker Smart
Apex2 diffractometer with Mo-K$_{\alpha}$ radiation, a graphite
monochromator with monocarp collimator, and a CCD area detector were
used for this experiment. The structure was refined with SHELXL-97
software using 1033 measured reflections of which 115 were unique
and 108 observed. The final residuals were $R_1$= 1.36\% and 1.96\%
for the observed data and $wR_2$= 3.31\% and 4.52\% for all data for
\Sr\ and \SrCacc\, respectively. Sr and Ca atoms were found to
reside in the same site with a refined Ca:Sr ratio of
0.33(1):0.67(1), giving the exact formula \SrCaxray\ from x-ray
analysis.

Figure~\ref{fig3} shows the unit cell of \SrCaxray\ determined by
single crystal x-ray diffraction at 250~K. The size of the
ellipsoids map the thermal agitation of the particular ion at 250~K
with a 50\% probability factor, which means the probability of
finding the center of the atom inside the ellipsoid.

Figure~\ref{fig4}presents the variation of the $a$- and $c$-axis
lattice constants (upper panel), the unit cell volume and the
tetragonal $c/a$ ratio (lower panel) with Ba-Sr and Sr-Ca
concentrations determined from refinements of the single crystal
x-ray diffraction data for \BaSr\ and \SrCa\ crystals taken at 250
K. Within experimental accuracy, the $a$- and $c$-axis lattice
constants, the $c/a$ ratio, and the unit cell volume all show a
monotonic linear decrease with alkaline earth substitution in the
continuous series from \Ba\ to \Sr\ to \Ca. This fact indicates that
the whole \BaSrCa\ series progression experiences a uniform chemical
pressure effect due to the reduction of the cation size that follows
Vegard's law, as expected for the decreasing ionic radii of Ba, Sr
and Ca, respectively.

The lattice parameters of \BaSr\ obtained in our experiments are
consistent with the data reported in a recent
study~\cite{Wang,Rotter2}, which found a systematic increase of
$T_0$ with increasing Sr content and no superconductivity. On the
other hand, substitution of arsenic for the smaller phosphorus
atoms, also instituting a chemical pressure effect, induces
superconductivity in \BaP~\cite{Wang}. Thus, a pressure-volume
effect is clearly an oversimplified explanation for
superconductivity in \BaP. In the future, it will be interesting to
investigate the evolution of superconductivity combining both the
chemical pressure effect of alkaline earth substitution studied here
and another tuning parameter that induces superconductivity in order
to investigate the role of lattice density in these phenomena.


In summary, we have systematically studied the crystallographic
properties in the solid solution series \BaSr\ and \SrCa\ by growing
high-quality single crystals by the self-flux method. X-ray
diffraction and data refinement reveal that the entire \BaSrCa\
series progresses according to Vegard's law, experiencing a uniform
chemical pressure effect due to substitution with reduced ionic size
from Ba to Sr to Ca, respectively.


\section{Acknowledgments}

\medskip
The authors acknowledge B.~W.~Eichhorn for experimental assistance,
and N.P.B. acknowledges support from a CNAM Glover fellowship. This
work was supported by AFOSR-MURI Grant FA9550-09-1-0603.





\smallskip

\end{document}